\documentclass[5p,times,merge,number,sort&compress]{elsarticle}

\usepackage[T1]{fontenc}
\usepackage[utf8]{inputenc}
\usepackage[english]{babel}

\usepackage{graphicx}
\usepackage{isotope}
\usepackage{physics}
\usepackage{xcolor}
\usepackage{subcaption}
\usepackage{hyperref}
\usepackage{amsmath}
\usepackage{multirow}
\usepackage{wasysym}
\usepackage{fdsymbol}
\usepackage{xspace}
\usepackage[nameinlink]{cleveref}

\crefname{table}{Tab.}{Tabs.}
\Crefname{table}{Table}{Tables}
\crefname{figure}{Fig.}{Figs.}
\Crefname{figure}{Figure}{Figures}
\crefname{equation}{Eq.}{Eqs.}
\Crefname{equation}{Equation}{Equations}

\DeclareMathAlphabet{\curly}{OMS}{cmsy}{m}{n}

\newcommand*{\elem}[2]{\ensuremath{\isotope[#2]{\mathrm{#1}}}}
\newcommand*{\lelem}[2]{\ensuremath{\isotope[#2\,][\Lambda]{\mathrm{#1}}}}

\newcommand*{\hw}{\ensuremath{\hbar\Omega}\xspace}
\newcommand*{\Nmax}{\ensuremath{N_\mathrm{max}}\xspace}
\newcommand*{\orb}[3]{\ensuremath{#1\mathrm{#2}_{#3/2}}\xspace}

\definecolor{MPLblue}{HTML}{1F77B4}
\definecolor{MPLorange}{HTML}{FF7F0E}
\definecolor{MPLgreen}{HTML}{2CA02C}
\definecolor{MPLred}{HTML}{D62728}
\definecolor{MPLpurple}{HTML}{9467BD}
\definecolor{MPLbrown}{HTML}{8C564B}
\definecolor{MPLpink}{HTML}{E377C2}
\definecolor{MPLgray}{HTML}{7F7F7F}
\definecolor{MPLolive}{HTML}{BCBD22}
\definecolor{MPLcyan}{HTML}{17BECF}

\begin{document}

\title{Halo Structures in p-Shell Hypernuclei with Natural Orbitals}

\address[tud]{Institut f\"ur Kernphysik, Fachbereich Physik, Technische Universit\"at Darmstadt, Schlossgartenstr. 2, 64289 Darmstadt, Germany}
\address[hfhf]{Helmholtz Forschungsakademie Hessen f\"ur FAIR, GSI Helmholtzzentrum, 64289 Darmstadt, Germany}

\author[tud]{Marco~Kn\"oll}
\ead{marco.knoell@physik.tu-darmstadt.de}
\author[tud,hfhf]{Robert~Roth}
\ead{robert.roth@physik.tu-darmstadt.de}

\date{\today}

\begin{abstract}

\noindent We extend the concept of natural orbitals as an optimized single-particle basis for ab initio nuclear many-body calculations to hypernuclei and show that their superior properties, in particular accelerated convergence and independence of the underlying harmonic-oscillator frequency, can be directly transferred to the hypernuclear regime as demonstrated in no-core shell model calculations for selected p-shell hypernuclei.
Moreover, the radial single-particle wavefunctions associated with the natural-orbital basis yield important structural information with respect to the different particle species allowing us to identify a hyperon halo in \lelem{He}{5}.
We further explore nucleonic and hyperonic halo structures in $A=6$ and $A=7$ singly-strange hypernuclei based on one-body densities and point-particle radii obtained from no-core shell model calculations with realistic interactions from chiral effective field theory.

\end{abstract}

\maketitle

\section{Introduction}

In recent years, modern ab initio approaches such as Faddeev-Yakubovsky calculations \cite{Nogga2002hypernuclei,Nogga2014light}, Gaussian expansion methods \cite{Hiyama2009structure}, Quantum Monte-Carlo approaches \cite{Lonardoni2013effects,Lonardoni2013auxiliary}, nuclear lattice calculations \cite{Frame2020impurity,Hildenbrand2024towards}, and no-core shell model (NCSM) variants \cite{Wirth2014abinitio,Wirth2018hypernuclear,Wirth2018light,Le2020jacobi} have been extended to describe systems with strangeness.
In particular the latter have proven very powerful in applications across various s- and p-shell hypernuclei together with realistic nucleonic and hyperon-nucleon (YN) interactions from chiral effective field theory (EFT) \cite{Polinder2006hyperon,Haidenbauer2020hyperon,Haidenbauer2023hyperon,Knoell2023Hyperon}.
In the NCSM the quantum many-body problem is treated as a large-scale matrix eigenvalue problem for the Hamiltonian that is expanded in a many-body Slater-determinant basis, which is constructed from an underlying single-particle basis.
Originally, the NCSM is formulated in a harmonic oscillator (HO) basis and the model space is truncated with respect to the total number of HO excitation quanta.
While the simplicity of the method gives access to a wide range of observables, it is apparent that the incorrect asymptotic shape of the HO wavefunctions hinders the convergence of observables that are sensitive to their long-range behavior.
Conceptually, the NCSM, and all other configuration-interaction (CI) methods, are not limited to the HO basis but can employ any other complete single-particle basis.
Due to the completeness of the bases they all converge to the same result, however, they do so at drastically different rates.

The search for an optimized single-particle basis has led to the development of natural orbitals (NAT), which have successfully be applied in various many-body methods in quantum chemistry \cite{Bender1966Natural,Hay1973On,Siu1974Configuration} and nuclear structure physics \cite{Constantinou2017Natural,Tichai2019Natural,Hoppe2021Natural,Robin2021Entanglement}.
Their most prominent features are an accelerated convergence of the many-body calculation as well as an independence of the HO frequency.
Conceptually, NAT are defined as the eigenvectors of the one-body density matrix for a given system.
This density matrix can be obtained through Hartree-Fock many-body perturbation theory (HF-MBPT) \cite{Roth2006Hartree,Shavitt2009Many,Tichai2016Hartree} up to second order, which provides a computationally cheap yet good approximation to the exact one-body density.
Note that this method includes bulk properties into the one-body density matrix resulting in a more physical single-particle basis compared to the HO basis.

Since NAT are informed by an approximate many-body state of the system under investigation, the single-particle wavefunction carry important structural information about the respective hypernucleus.
This information can be accessed for the different particle species individually, thus, making NAT an ideal diagnostic tool to study halo structures in hypernuclei by looking at the spacial extend of the radial single-particle wavefunctions.
These structures are of particular interest as they give insights into the details of the hyperon-nucleon interaction.
Na{\"i}vely, hyperons in singly-strange hypernuclei are expected to populate the lowest-lying orbitals and, therefore, be localized in the interior of the nucleus, since they are not affected by the Pauli principle.
However, investigations with cluster models, EFTs, or Skyrme-Hartree-Fock approaches indicate hyperon halo structures in several light hypernuclei \cite{Hiyama1996Three,Cobis1997The,Hiyama2009structure,Hildenbrand2019Three,Zhang2021Hyperon}.
Moreover, the addition of the hyperon is expected to shift the proton or neutron drip line outwards due to the additional attraction between nucleons and hyperons, also facilitating the formation of nucleonic halo structures in hypernuclei.
Experimentally this has been found to be the case for e.g.\ \lelem{Be}{7}, which is more strongly bound than \lelem{He}{5}, while its nucleonic parent \elem{Be}{6} is beyond the proton drip line for $N=2$ located at \elem{He}{4} \cite{NUBASE2020,JURIC1973A}. 
Halo nuclei are difficult to describe within the HO basis since the core of the nucleus and its halo have different radial extends, while the HO basis comes with a single length scale that corresponds to the oscillator length. 
A single-particle basis that is optimized for different particle species would, therefore, greatly benefit the description of systems with halo structures.

In this work we investigate these phenomena in an ab initio framework.
Following the derivations in \cite{Tichai2019Natural} we extend the concept of NAT to hypernuclei by considering $\Lambda$ and $\Sigma$ hyperons alongside nucleons.
We then study their impact on NCSM calculations for selected p-shell hypernuclei with a focus on convergence rate and dependence on the HO frequency \hw.
Afterwards we investigate the spatial structures of \lelem{He}{5} and \lelem{Li}{7}, as well as the more neutron-rich \lelem{He}{6}, \lelem{He}{7} and proton-rich \lelem{Li}{6}, \lelem{Be}{7} hypernuclei by looking at one-body densities and point-particle radii for the individual particles species. 

\section{Formalism}

\paragraph{Hypernuclear Natural Orbitals}

The natural orbitals are defined as the eigenstates of the one-body density matrix
\begin{flalign}
\rho_{pq} = \mel*{\Psi}{c^\dagger_p c_q}{\Psi},
\end{flalign}
where, for our purpose, the many-body state $\ket*{\Psi}$ is approximated up to second order in HF-MBPT and $c^\dagger_p, c_q$ are creation and annihilation operators in an auxiliary basis.
Here, $p=\{n_p,l_p,j_p,m_{j_p},\chi_p\}$ is a collective index for the radial quantum number $n_p$, the orbital angular momentum $l_p$, the total angular momentum $j_p$ and its projection $m_{j_p}$, and particle species $\chi_p$.
We further impose spherical symmetry on the Hamiltonian resulting in a correlated density matrix that is block diagonal w.r.t. all quantum numbers except for $n$.
Hence, the particle species separate and the derivation of the perturbed density matrix appears identical to the purely nucleonic case.
Nevertheless, we will recall the most important formulas in order to point out where adjustments for hyperons enter.
This is already relevant in the single particle index $p$, where the particle species $\chi_p$ covers not only neutrons and protons but also the $\Lambda$ and $\Sigma$ hyperons.
Note that we limit ourselves to singly-strange hypernuclei and, therefore, $\Xi$ hyperons are omitted.

As already stated, we choose to approximate the reference state in HF-MBPT due to its computational efficiency.
For the same reason we only consider a single-determinant HF reference state.
Since singly-strange hypernuclei are by default open-shell systems, this state is constructed in an equal filling approximation introducing fractional occupation numbers in degenerate shells \cite{Perez2008Microscopic}.
We further limit the MBPT to a two-body Hamiltonian, however, three-body forces are taken into account via a normal-ordered two-body (NO2B) approximation \cite{Roth2012Medium,Binder2013Ab} w.r.t.\ the aforementioned single-determinant HF reference state.

The perturbed density matrix can then be expressed as
\begin{flalign}
    \rho_{pq} \approx \rho^\mathrm{HF}_{pq} + \rho^{(02)}_{pq} + \rho^{(20)}_{pq} + \rho^{(11)}_{pq},
\end{flalign}
where $\rho^\mathrm{HF}_{pq}$ is the unperturbed HF density and expressions for the corrections are given by
equations (4)--(7) in \cite{Tichai2019Natural}.
Again, we emphasize that the indices in these expressions also cover hyperon single-particle states.

Diagonalizing the correlated density matrix yields the NAT basis, which can be expressed as a superposition of HO basis states such that
\begin{flalign}
    \ket*{nljm_j\,\chi}_\mathrm{NAT} = \sum_{n'} c^{(lj\chi)}_{nn'} \ket*{n'ljm_j\,\chi}_\mathrm{HO} \label{eq:expansion}
\end{flalign}
with expansion coefficients $c^{(lj\chi)}_{nn'}$.
\begin{figure*}[t]
    \centering
    \includegraphics[width=\textwidth]{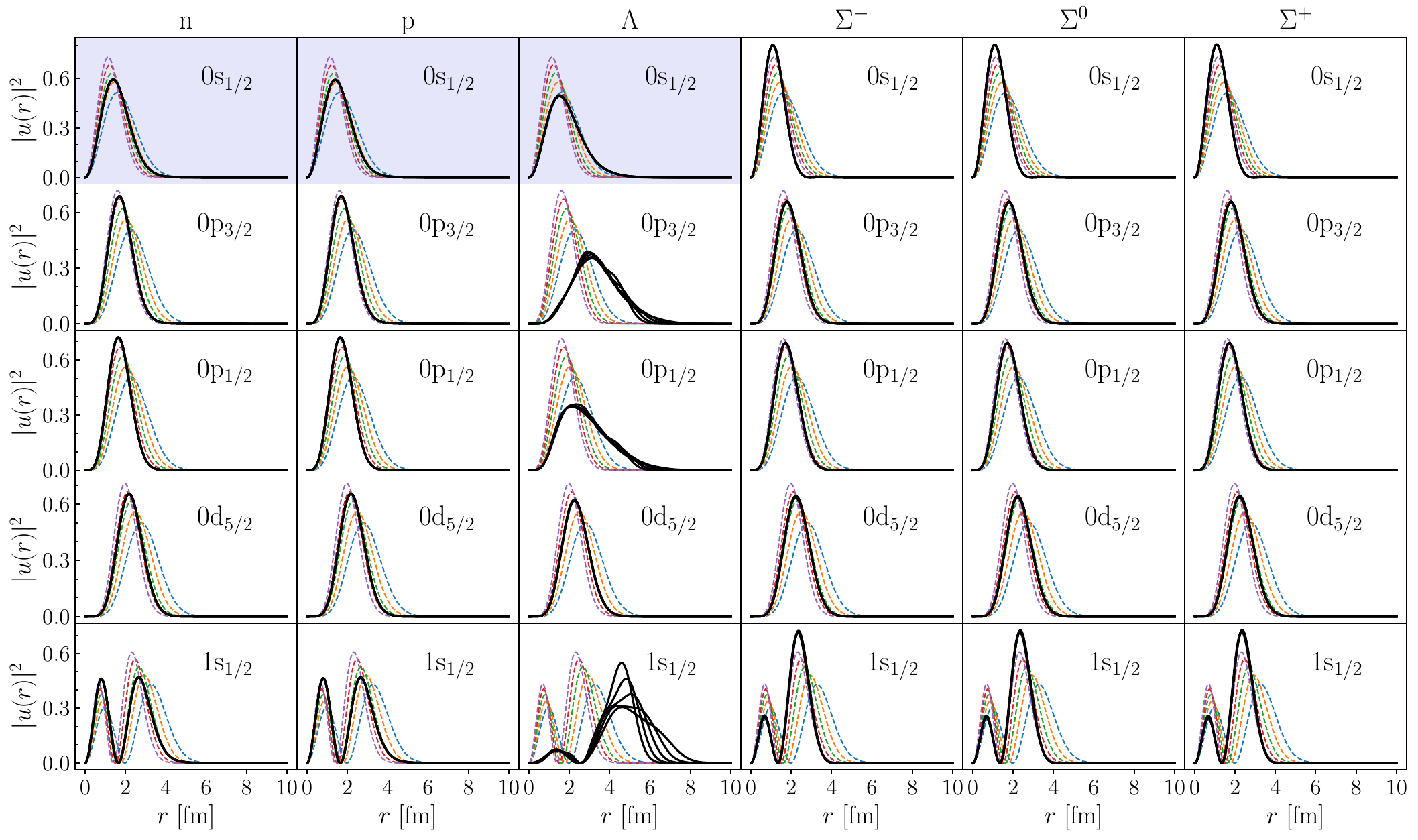}
    \caption{Squared single-particle radial wavefunctions in units of $\mathrm{fm}^{-1}$ in HO basis (colored) and NAT basis for \lelem{He}{5} (black) for $\hbar\Omega=16~(\textcolor{MPLblue}{\textbf{---}}),20~(\textcolor{MPLorange}{\textbf{---}}),24~(\textcolor{MPLgreen}{\textbf{---}}),28~(\textcolor{MPLred}{\textbf{---}}),$ and $32~(\textcolor{MPLpurple}{\textbf{---}})$~MeV.
    Shown here are the five lowest-lying orbitals (rows) for all considered particle species (columns).
    Colored background indicates occupied orbitals.}
    \label{fig:LHe5_wfct_full}
\end{figure*}

According to \cref{eq:expansion} we can transform the matrix elements for the hypernuclear Hamiltonian and other operators to the NAT basis and employ them in subsequent many-body calculations.
Note, that we include a term in the Hamiltonian that accounts for the different hyperon and nucleon masses.
For all applications discussed in this work we use non-local chiral nucleon-nucleon (NN) and three-nucleon (3N) interactions at next-to-next-to-next-to-leading order (N$^3$LO) with cutoff $\Lambda_\mathrm{N}=500$~MeV from \cite{Huether2020family} accompanied by a non-local chiral YN interaction at LO with cutoff $\Lambda_\mathrm{Y}=700$~MeV from \cite{Knoell2023Hyperon}.
We, further, apply a similarity renormalization group (SRG) transformation \cite{Wirth2016induced,Wirth2019similarity}.
All interactions are consistently evolved up to flow parameter $\alpha=0.08$~fm$^4$ and induced 3N and hyperon-nucleon-nucleon (YNN) forces are taken into account. 
Finally, we employ an NO2B approximation, which significantly reduces the computational cost of the matrix element transformation at the price of an error in the order of 1\% \cite{Tichai2019Natural,Binder2013Ab}.
The HO basis and, thus, the resulting NAT basis is truncated at $e_\mathrm{max}=12$ with an additional truncation on the orbital angular momentum $l_\mathrm{max}=8$.
The initial matrix elements for the 3N and YNN interactions are further truncated by an upper limit to the sum over the three principal quantum numbers given by $E^\mathrm{3N}_\mathrm{3max}=14$ and $E^\mathrm{YNN}_\mathrm{3max}=12$ respectively.

\section{Properties of the Natural Orbitals}

\paragraph{Single-Particle Wavefunctions}

We begin our investigation of the hypernuclear natural orbitals by comparing them to the HO basis in, both, the single-particle wavefunctions and the many-body convergence.
Starting with the former, \cref{fig:LHe5_wfct_full} shows the squared single-particle radial wavefunctions for the HO basis (colored) and the NAT basis for \lelem{He}{5} (black) for all active particle species in the five lowest-lying orbitals for a range of \hw.
We immediately see that in almost all cases the NAT wavefunctions are independent of \hw.
This is in strong contrast to the HO wavefunctions.
However, there are some cases in which the NAT wavefunctions retain a frequency dependence.
This is most pronounced in the \orb{1}{s}{1} orbital for the $\Lambda$ hyperon but also visible in the corresponding \orb{0}{p}{3} and \orb{0}{p}{1} orbitals. 
Apparently, the remaining frequency dependence only occurs for spatially extended orbitals with significant long-range components.
Modeling such a long-range behavior in a HO basis is difficult as it requires either very small values of \hw that correspond to large oscillator lengths, which directly translate to the spatial extend of the radial wavefunction, or large radial quantum numbers $n$.
As our HO basis is truncated at $e_\mathrm{max}=12$ we are limited to $n\leq6$.
Hence, for larger \hw the HO basis, which the NAT basis is constructed from, does not contain sufficiently extended states to model the NAT wavefunctions.
Consequently, a residual \hw dependence appears for spatially extended NAT states as an artifact of the HO truncation.

Comparing the different particle species we find that the wavefunctions for neutrons and protons are almost identical.
This is to be expected as \lelem{He}{5} has equal numbers of both nucleon species.
A similar reasoning applies to the NAT for the $\Sigma$ hyperons, which are again indistinguishable for the different particles and contribute to the hypernucleus in equal average particle numbers of $\mathrm{n}_\Sigma\approx0.002$.
Due to this small contribution we expect the $\Sigma$ orbitals to be predominantly constrained by the orthogonality condition on the basis.
We will, therefore, only consider neutron, proton, and $\Lambda$ orbitals in the following.

\paragraph{Convergence in the No-Core Shell Model}

The other aspect of major interest is how the NAT impact many-body calculations.
We investigate this in the importance truncated (IT) NCSM \cite{Roth2009importance,Wirth2018hypernuclear} for which the hypernuclear many-body problem is cast into a matrix eigenvalue problem
\begin{align}
    \sum_j\mel*{\phi_i}{H}{\phi_j}\braket*{\phi_j}{\Psi_n} = E_n\braket{\phi_i}{\Psi_n} \quad \forall i
\end{align}
through an expansion in Slater determinants $\ket{\phi_i}$ with Hamiltonian $H$, energy eigenvalues $E_n$, and corresponding eigenstates $\ket{\Psi_n}$.
The resulting matrix eigenvalue problem is then truncated w.r.t.\ \Nmax.
We transfer this truncation scheme to the NAT basis w.r.t.\ the new radial quantum number $n'$.

Slater determinants can be constructed from any complete single-particle basis.
However, the separation of intrinsic and center-of-mass components is guaranteed only in the HO basis.
Hence, many-body states obtained in NAT basis will contain residual center-of-mass contributions.
We can shift those to higher-lying parts of the spectrum by modifying the Hamiltonian with a Lawson term.
The strength of this term is controlled by a prefactor $\lambda_\mathrm{cm}$ and we monitor its expectation value to ensure it remains small.
For all applications here $\lambda_\mathrm{cm}=0.3$ is used and the expectation value varies between 0.1 and 0.2~MeV across the different hypernuclei and HO frequencies at the highest \Nmax respectively.

\begin{figure}
    \centering
    \includegraphics[width=\columnwidth]{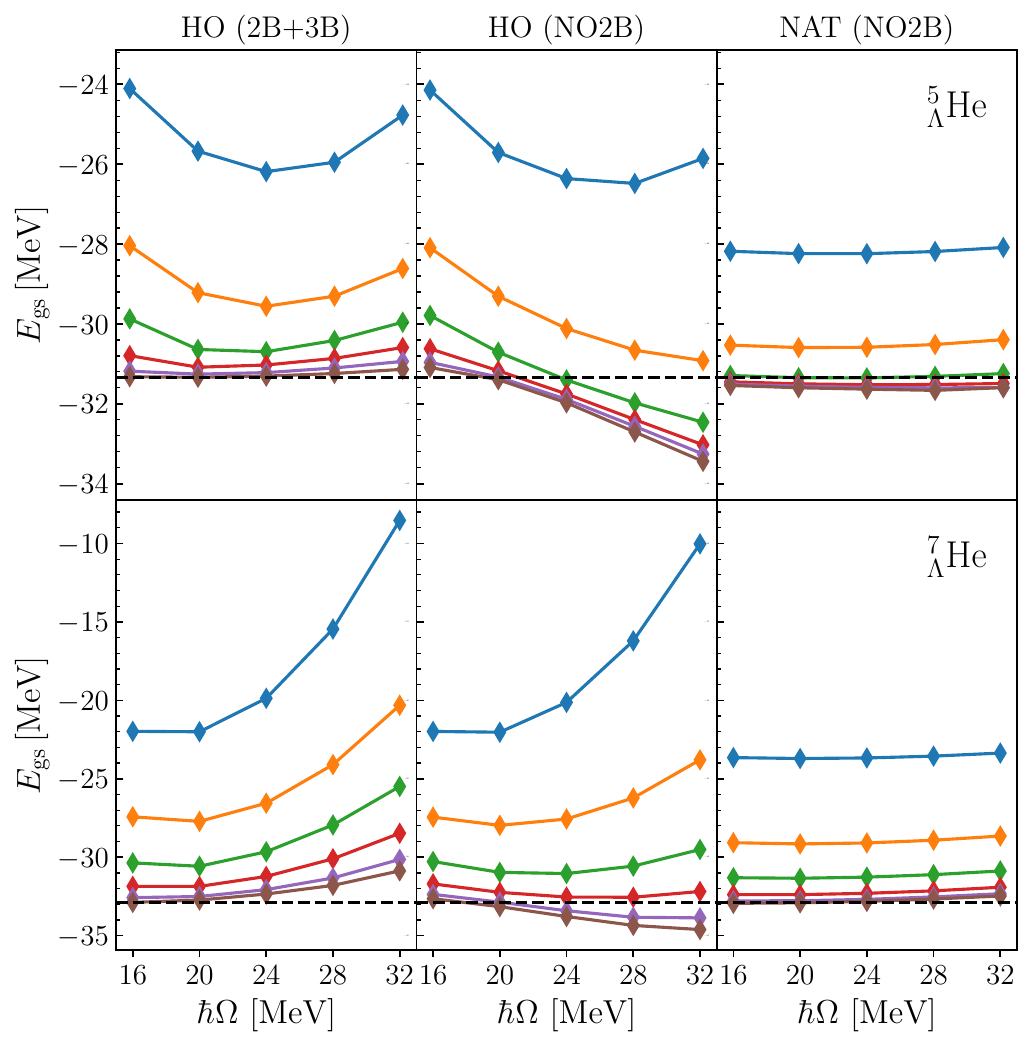}
    \caption{NCSM calculations of ground-state energies of \lelem{He}{5} (upper panels) and \lelem{He}{7} (lower panels) for HO basis with full two-body and three-body forces (left), HO basis with NO2B approximation (center), and NAT basis (right) as a function of \hw.
    Different colors correspond to $\Nmax=2$~(\textcolor{MPLblue}{\textbf{---}}), 4~(\textcolor{MPLorange}{\textbf{---}}), 6~(\textcolor{MPLgreen}{\textbf{---}}), 8~(\textcolor{MPLred}{\textbf{---}}), 10~(\textcolor{MPLpurple}{\textbf{---}}), and 12~(\textcolor{MPLbrown}{\textbf{---}}). The black dashed lines indicate the lowest energy from the full HO calculation for comparison.
    }
    \label{fig:NCSM_compare_bases}
\end{figure}
In \cref{fig:NCSM_compare_bases} NCSM calculations of the ground-state energy of \lelem{He}{5} and \lelem{He}{7} are shown in HO basis with full two-body and three-body forces, in HO basis with NO2B approximation, and in NAT basis as a function of \hw.
For the HO basis in the left and center panels we see the expected frequency dependence for both hypernuclei. 
We can further assess the effects of the NO2B approximation and find that for small \hw the NO2B error is very small, while it increases significantly for larger \hw. 
This effect is more pronounced in \lelem{He}{5} than in \lelem{He}{7}.
Moving on to the NAT basis shown in the right panels we find the anticipated frequency independence we have already seen for the radial wavefunctions. and, despite using the NO2B approximation, the converged values are in good agreement with the results for the HO basis with three-body forces. 
Moreover, the convergence rate is significantly increased compared to the calculations in HO basis.
This improvement is more significant than for nucleonic systems, where the convergence rate in the NAT basis is similar to the optimal \hw in HO basis \cite{Tichai2019Natural}.

\begin{figure}
    \centering
    \includegraphics[width=.95\columnwidth]{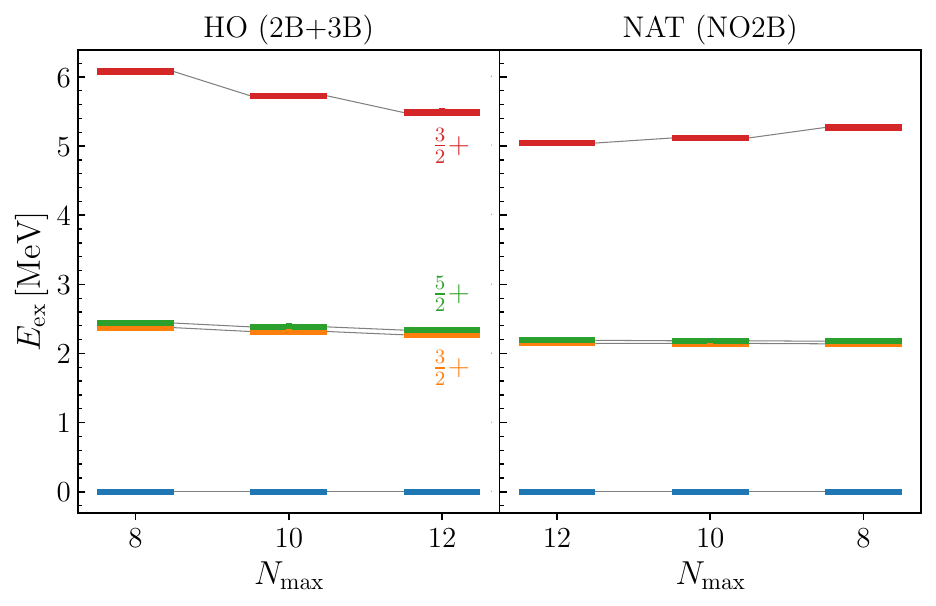}
    \caption{Excitation energies of the lowest-lying natural-parity states of \lelem{He}{7} in HO basis (left) and NAT basis (right) for $\Nmax=8,10,$ and $12$ at $\hw=16$~MeV.}
    \label{fig:LHe7_compare_spectra}
\end{figure}
We can further look at the spectra of hypernuclei.
\Cref{fig:LHe7_compare_spectra} shows the excitation energies of the three lowest-lying excited natural parity states in \lelem{He}{7} for the HO and NAT bases.
Across all three states we find good agreement between the bases and the error of the NO2B approximation is negligible.

\paragraph{One-Body Densities and Radii}

In order to study the structure of hypernuclei we consider particle-specific one-body densities and point-particle radii calculated from the eigenstates obtained within the NCSM calculation $\ket{\Psi_\mathrm{NCSM}}$.
For a particle species $\chi$ the normalized one-body density is given by
\begin{flalign}
    \rho_\chi (r) = \frac{1}{4\pi\expval{n_\chi}} \sum_{pq} P^\chi_p P^\chi_q \, \rho^\mathrm{HO}_{pq}\, R_{n_pl_p}(r) R_{n_ql_q}(r)\delta^{l_p}_{l_q}\delta^{m_{l_p}}_{m_{l_q}},
\end{flalign}
with HO radial wave function $R_{nl}$, particle number and projection operators
\begin{flalign}
    n_\chi = \sum^A_{i=1}  P^{\chi}_p, \quad
    P^{\chi}_p = \begin{cases}
        1 & \text{if particle $p$ is of species $\chi$,}\\
        0 & \mathrm{else},
    \end{cases}
\end{flalign}
and the one-body density matrix in HO basis $\rho^\mathrm{HO}$, which is obtained from the one in NAT basis using \cref{eq:expansion}
\begin{flalign}
    \rho^\mathrm{HO}_{pq} = \sum_{p',q'} c^{(l_pj_p\chi_p)}_{n_pn_{p'}} c^{(l_qj_q\chi_q)}_{n_qn_{q'}} \mel*{\Psi_\mathrm{NCSM}}{c^\dagger_{p'} c_{q'}}{\Psi_\mathrm{NCSM}}
\end{flalign}
For brevity the decoupling of the angular momenta in the single-particle states has been omitted.

Further, we can calculate root-mean-square (rms) mass radii along with point-particle mean-square radii for which the corresponding operator in a translationary invariant form is given by
\begin{flalign}
    R^2_{\chi,\mathrm{ms}} = \frac{1}{M\expval{n_\chi}} \sum_{p<q} \qty(m_pP^{\chi}_q + m_qP^{\chi}_p - n_\chi\frac{m_pm_q}{M})r^2_{pq}
\end{flalign}
with single-particle mass operator $m_p$ and total mass operator $M$ \cite{Wirth2018diss}.

\begin{figure}[t]
    \centering
    \includegraphics[width=.99\columnwidth]{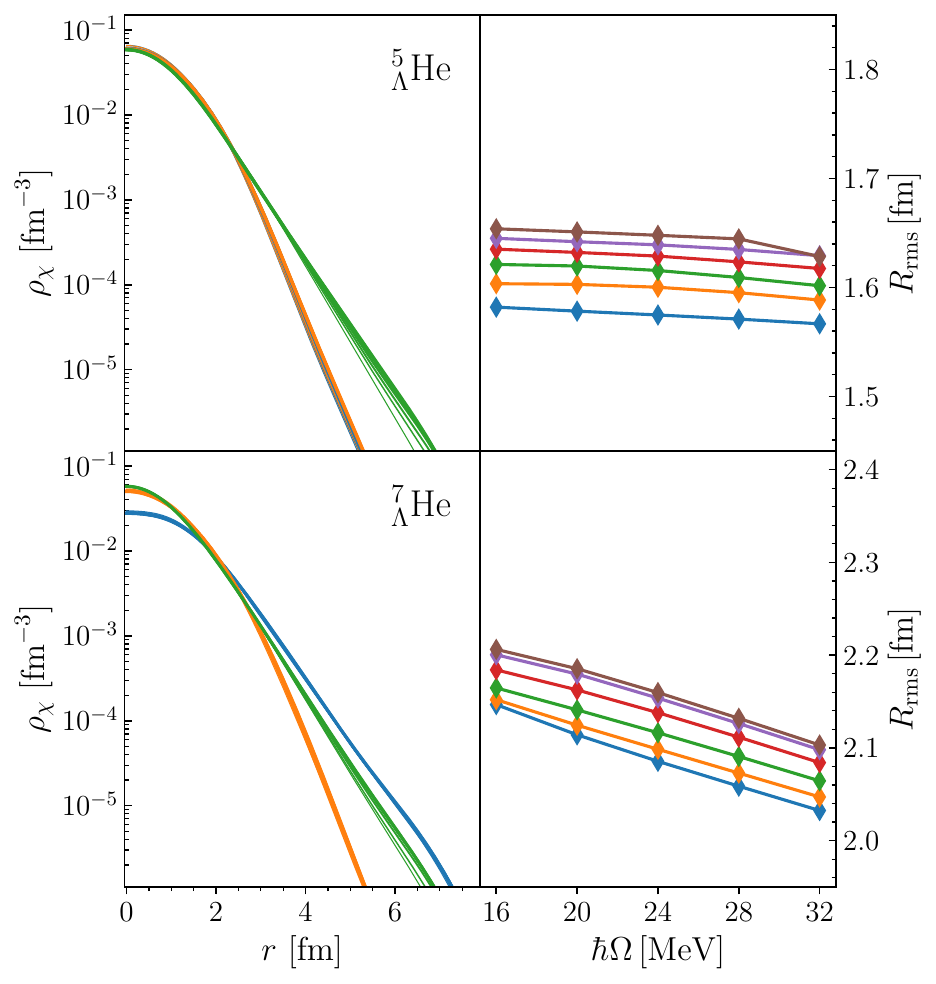}
    \caption{(Left-hand panels) Particle-specific one-body densities in \lelem{He}{5} and \lelem{He}{7} for neutrons~(\textcolor{MPLblue}{\textbf{---}}), protons~(\textcolor{MPLorange}{\textbf{---}}), and $\Lambda$ particles~(\textcolor{MPLgreen}{\textbf{---}}) up to $\Nmax=14,12$ for $\hw=16$~MeV. 
    Line thickness increases with model-space size. 
    (Right-hand panels) Rms mass radii for \lelem{He}{5} and \lelem{He}{7} as a function of \hw for $\Nmax=2$~(\textcolor{MPLblue}{\textbf{---}}), 4~(\textcolor{MPLorange}{\textbf{---}}), 6~(\textcolor{MPLgreen}{\textbf{---}}), 8~(\textcolor{MPLred}{\textbf{---}}), 10~(\textcolor{MPLpurple}{\textbf{---}}), and 12~(\textcolor{MPLbrown}{\textbf{---}}).}
    \label{fig:OBD_radii}
\end{figure}
The left-hand panels of \cref{fig:OBD_radii} show the convergence of the neutron, proton, and $\Lambda$ densities w.r.t.\ \Nmax in \lelem{He}{5} and \lelem{He}{7}.
For both isotopes we find rapid convergence of the proton and neutron densities, while the $\Lambda$ density converges at a slower rate.
In both hypernuclei we see indications of halo structures in the hyperon density and for \lelem{He}{7} also in the neutron density, where the latter is to be expected as its parent nucleus \elem{He}{6} is a well-known neutron halo nucleus. 
We will discuss this more in detail in the next section.

Studying the spatially extended densities more closely, we find that they seem to bent for larger $r$, deviating from the exponential behavior.
Since the densities are converged w.r.t.\ \Nmax this, again, is an artifact from the $e_\mathrm{max}$ truncation of the underlying HO basis as we have already seen in the NAT wavefunctions.
This translates to the mass radii shown in the right-hand panels of \cref{fig:OBD_radii}.
While they are nearly independent of \hw in \lelem{He}{5} we find a systematic frequency dependence in \lelem{He}{7}.
Again, the calculated radii are nearly converged w.r.t.\ \Nmax.
Yet, they are sensitive to the tails of the densities and, therefore, affected by the $e_\mathrm{max}$ truncation.
This is more pronounced in systems with halo-structures in which the truncation error gets larger with increasing \hw.
\begin{figure}[t]
    \centering
    \includegraphics[width=\columnwidth]{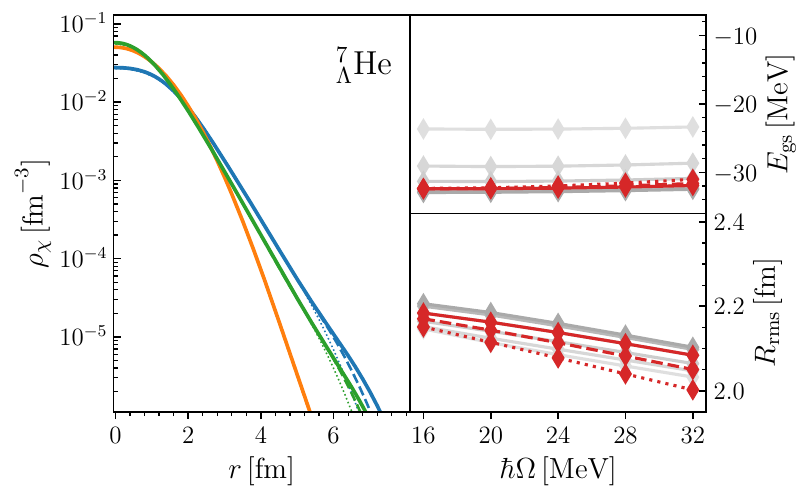}
    \caption{(Left-hand panel) Particle-specific one-body densities in \lelem{He}{7} for neutrons~(\textcolor{MPLblue}{\textbf{---}}), protons~(\textcolor{MPLorange}{\textbf{---}}), and $\Lambda$ particles~(\textcolor{MPLgreen}{\textbf{---}}) for $\Nmax=8$ and $\hw=16$~MeV.
    (Right-hand panels) Ground-state energies (upper) and rms mass radii (lower) for \lelem{He}{7} as a function of \hw for $\Nmax=8$ (red).
    Gray markers indicate data for $\Nmax=2$ to $12$ from \cref{fig:NCSM_compare_bases,fig:OBD_radii} for comparison.
    All results are shown for $e_\mathrm{max}=8$ (dotted), 10 (dashed), and 12 (solid).}
    \label{fig:emax_conv}
\end{figure}
Some exploratory calculations with larger $e_\mathrm{max}$ indicate that complete frequency independence is only achieved beyond truncations that are computationally feasible for now.
Nevertheless, we can investigate the $e_\mathrm{max}$ dependence in the accessible range as shown for \lelem{He}{7} in \cref{fig:emax_conv}.
We find that the energies are largely converged w.r.t.\ $e_\mathrm{max}$.
However, the tails of the spatially extended densities and the corresponding radii show the expected dependence on the $e_\mathrm{max}$ truncation comparable to the \Nmax truncation.
These deficiencies in the long-range behavior of the NAT wavefunctions can be reduced through the choice of a wider HO basis, which corresponds to smaller values of \hw. 
In fact, we find that the $e_\mathrm{max}$ truncation effects are less pronounced for small \hw.
Therefore, we will only consider calculations for $\hw=16$~MeV in the results section.
\begin{figure*}
    \centering%
    \subfloat[]{
        \includegraphics[width=0.46\textwidth]{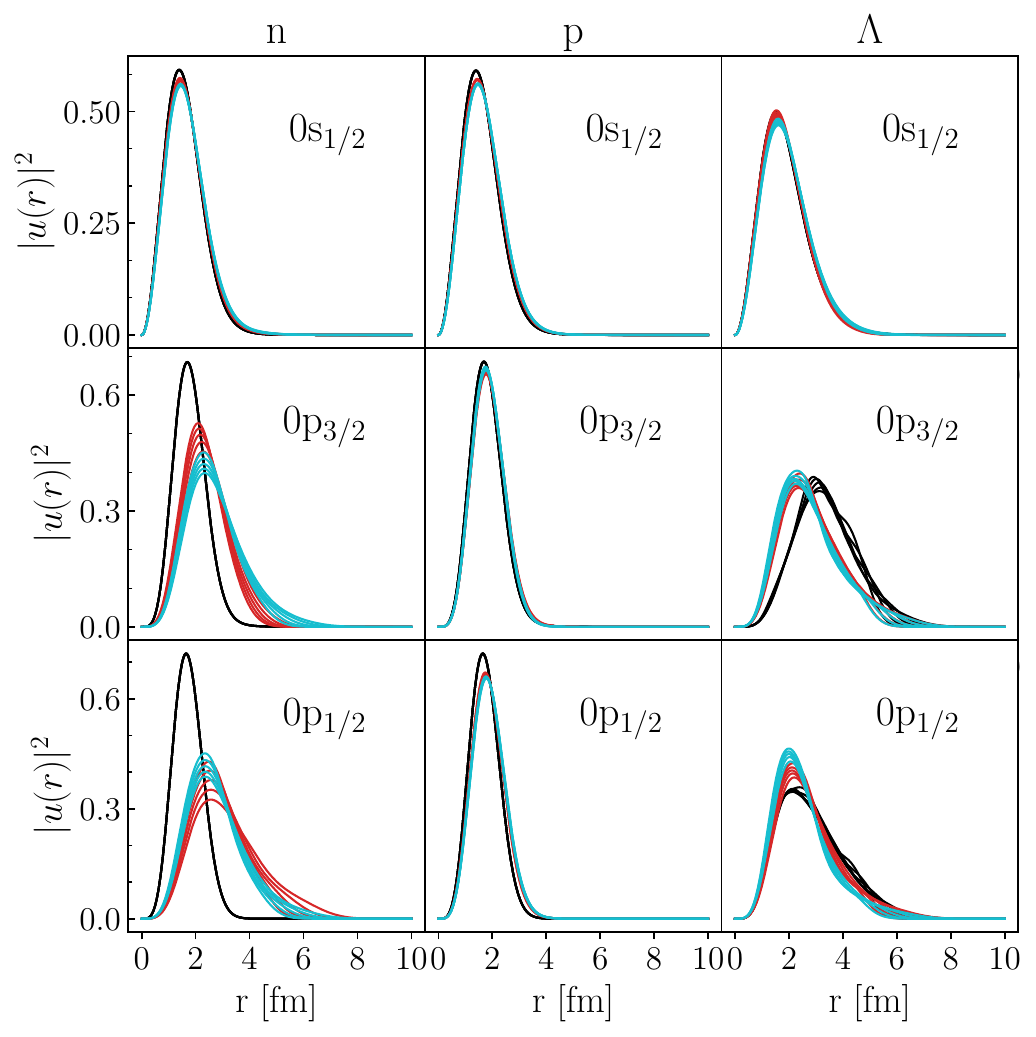}
        \label{fig:neutron_shift}}
    \qquad
    \subfloat[]{
        \includegraphics[width=0.46\textwidth]{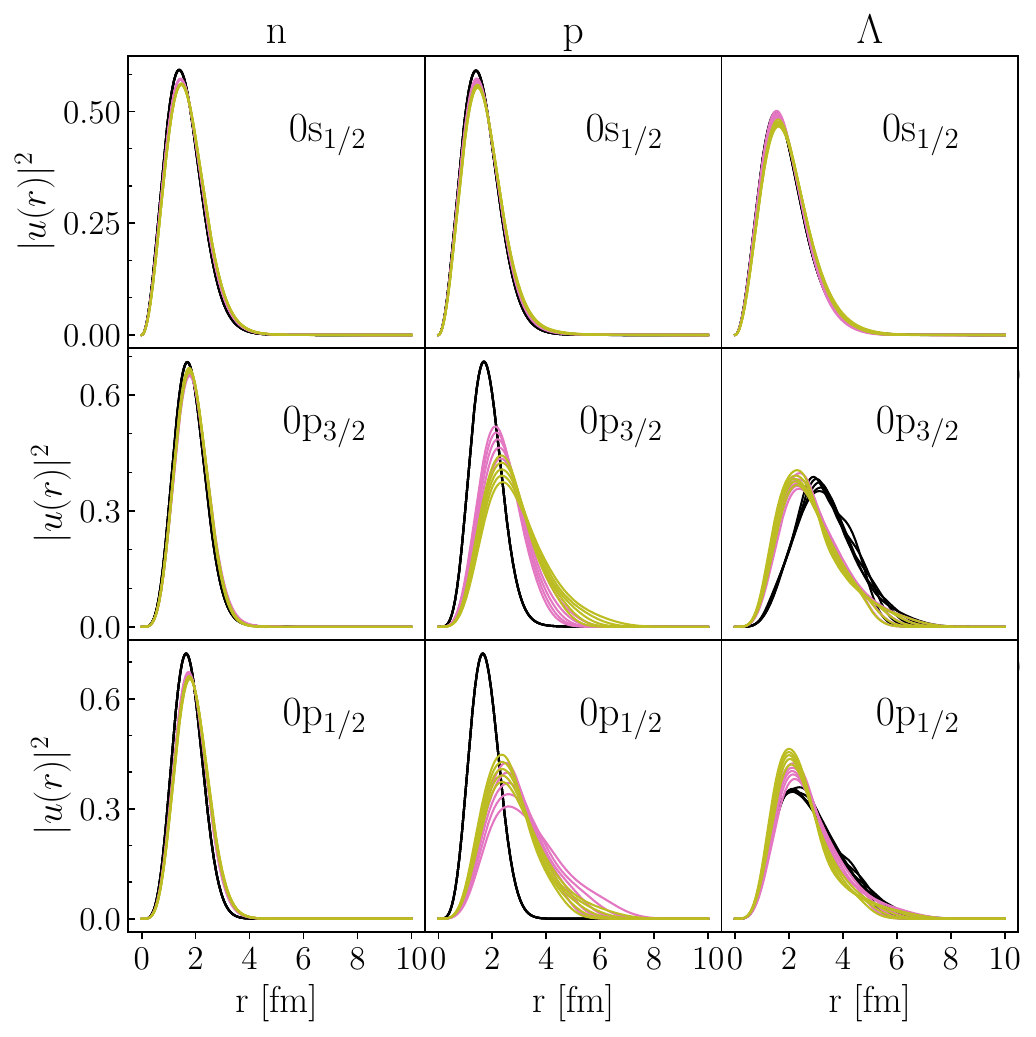}
        \label{fig:proton_shift}
    }
    \caption{
    Squared radial wavefunction in units of $\mathrm{fm}^{-1}$ in NAT basis for neutrons, protons, and $\Lambda$ hyperons in the three lowest-lying orbitals for $\hw=16,20,14,28,$ and $32$~MeV. (a) \lelem{He}{5}~(\textcolor{black}{\textbf{---}}) and the neutron-rich isotopes \lelem{He}{6}~(\textcolor{MPLred}{\textbf{---}}) and \lelem{He}{7}~(\textcolor{MPLcyan}{\textbf{---}}). (b) \lelem{He}{5}~(\textcolor{black}{\textbf{---}}) and the proton-rich isotones \lelem{Li}{6}~(\textcolor{MPLpink}{\textbf{---}}) and \lelem{Be}{7}~(\textcolor{MPLolive}{\textbf{---}}).
    }
    \label{fig:nucleon_shift}
\end{figure*}

\paragraph{Negative Occupation Numbers}

In order to be physically meaningful, the NAT one-body density must be a positive-definite operator featuring only non-negative eigenvalues smaller than or equal to one. 
But, this is not guaranteed to be the case in the framework discussed here and the occurrence of negative occupation numbers in open-shell systems has repeatedly been reported in both nuclear and molecular systems \cite{Hoppe2021Natural,Gordon1999A}.
Therefore, it is no surprise that we also encounter these issues in some hypernuclei.
They are particularly pronounced in triply open-shell systems such as \lelem{Li}{7}.
While small negative occupation numbers typically show no or negligible effects in many-body calculations, we encounter occupation numbers of $-0.6$ and below, depending on \hw, which manifest as pathological behavior in the NAT wavefunctions and the subsequent many-body calculations.
This slows down the model-space convergence, however, these effects are mitigated with increasing \Nmax.

As discussed in \cite{Gordon1999A}, the occurrence of negative occupation numbers indicates a breakdown of the single-reference ansatz for the density matrix caused by the combination of equal filling approximation and MBPT .
Hence, multi-reference approaches could improve the quality of the NAT basis.

\section{Results}

After discussing the properties of the NAT basis and its impact on many-body calculations, we turn to an investigation of \lelem{He}{5} and its $Z=2$ isotopes and $N=2$ isotones in order to study the structural effects emerging with the addition of nucleons to the system.
We start our investigation with a detailed discussion of the NAT wavefunctions depicted in \cref{fig:nucleon_shift} for neutrons, protons, and $\Lambda$ hyperons in the three lowest orbitals, of which the first two are (partially) occupied. 
In order to detect remaining \hw dependencies multiple frequencies are included.

Before we consider additional nucleons let us first discuss \lelem{He}{5} shown in black.
We find that the $\Lambda$ wavefunctions extend considerably further than the nucleonic wavefunctions across all orbitals, which hints at a hyperon halo.
Given that the neutron separation energy $S_n = 20.578$~MeV \cite{NUBASE2020} is much larger than the hyperon separation energy $B_\Lambda = 3.12(2)$~MeV \cite{Davis2005years} it seems likely that \lelem{He}{5} can be understood as a hyperon coupled to an $\alpha$ particle.

Similar structures can be found in the neutron-rich isotopes \lelem{He}{6} and \lelem{He}{7} depicted in \cref{fig:neutron_shift}.
While the wavefunctions of the \orb{0}{s}{1} orbital remains unchanged for all particle species, the neutron wavefunctions of the p orbitals shift to larger $r$ and show signs of \hw dependence. 
This shift is slightly more pronounced in \lelem{He}{7}.
Moreover, we see that the $\Lambda$ wavefunctions in the \orb{0}{p}{3} experience a shift towards a smaller range.
In a many-body context, we can further discuss the NAT occupation numbers for additional insights.
Outside of the \orb{0}{s}{1} orbital, which is almost fully occupied in both nuclei, the majority of the additional neutron in \lelem{He}{6} occupies the \orb{1}{p}{3} orbital with $n_\mathrm{n}\approx 0.66$ to $0.81$ depending on the frequency, which changes to $n_\mathrm{n}\approx 1.69$ to $1.76$ in \lelem{He}{7}.
Hence, both additional neutrons occupy roughly the same orbitals.

\begin{figure}
    \centering
    \includegraphics[width=1\columnwidth]{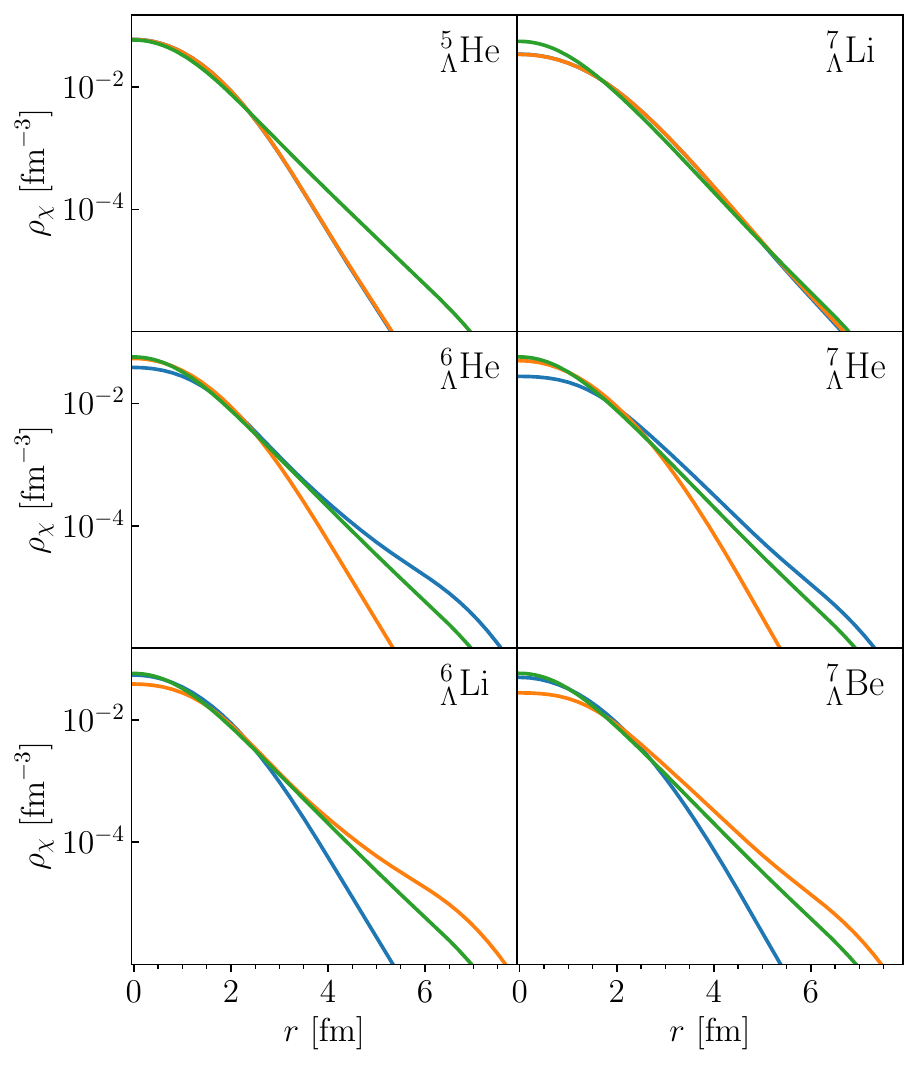}
    \caption{One-body densities for neutrons~(\textcolor{MPLblue}{\textbf{---}}), protons~(\textcolor{MPLorange}{\textbf{---}}), and $\Lambda$ hyperons ~(\textcolor{MPLgreen}{\textbf{---}}) obtained from IT-NCSM calculations at $\Nmax=14$ for \lelem{He}{5}, \lelem{He}{6}, and \lelem{Li}{6} and at $\Nmax=12$ for \lelem{Li}{7}, \lelem{He}{7}, and \lelem{Be}{7} at $\hw=16$~MeV.}
    \label{fig:all_densities}
\end{figure}
\begin{figure}
    \centering
    \includegraphics[width=1\columnwidth]{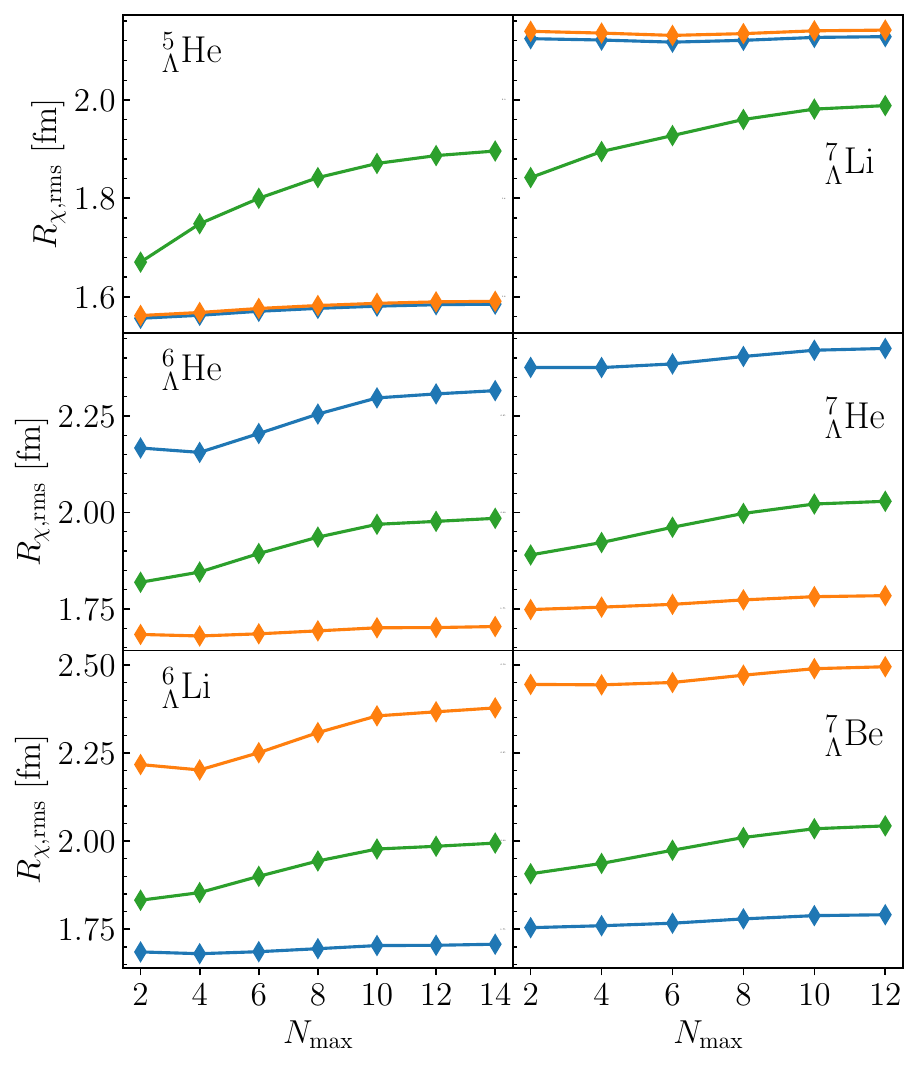}
    \caption{Point-particle rms radii for neutrons~(\textcolor{MPLblue}{\textbf{---}}), protons~(\textcolor{MPLorange}{\textbf{---}}), and $\Lambda$ hyperons ~(\textcolor{MPLgreen}{\textbf{---}}) in different p-shell hypernuclei for $\hw=16$~MeV.}
    \label{fig:all_radii}
\end{figure}
Comparing this to the results for \lelem{Li}{6} and \lelem{Be}{7} shown in \cref{fig:proton_shift} we find the analogous behavior but mirrored from neutrons to protons. 
The proton wavefunctions are shifted in the same manner and the $\Lambda$ wavefunctions behave identical to the previous case, agnostic to the type of nucleon added. 

We can investigate potential halo structures more closely through the particle-specific one-body densities in \cref{fig:all_densities} together with the corresponding point-particle radii depicted in \cref{fig:all_radii}.
Starting once more with \lelem{He}{5} we find a clear distinction between the nucleonic densities, which are indistinguishable, and the $\Lambda$ density that extends far beyond the nucleonic ones. 
Consequently, the $\Lambda$ radius is much larger than the nucleonic radii, which only differ by a minor shift caused by the Coulomb interaction.
Moving on to the heavier He isotopes we find the changes in the wavefunctions confirmed.
The proton and $\Lambda$ densities remain unchanged, while the neutron density exhibits strong long-range components that are even more pronounced in \lelem{He}{6} than in \lelem{He}{7}.
We, again, find that the tails of the neutron densities are bent indicating incomplete convergence w.r.t.\ $e_\mathrm{max}$.
We obtain proton and $\Lambda$ radii that slightly increase with particle number and neutron radii that extend beyond the proton radii.
We recall that the neutron radii in particular are underestimated due to truncation effects.
Hence, we find a two-layered halo structure with an $\alpha$ core, a hyperon halo, and beyond that a neutron halo.
Judging from the densities, the basis truncation might result in a larger error for \lelem{He}{6} than for \lelem{He}{7} impeding a well-founded comparison of the two.
For their mirror nuclei \lelem{Li}{6} and \lelem{Be}{7} we, again, find similar results with neutrons and protons exchanged for, both, densities and radii with the exception of slightly stronger long-range components and slightly larger radii due to Coulomb effects.

Finally, we can look at our diagnostics for \lelem{Li}{7}.
Besides the deficits caused by the negative occupation numbers, we can qualitatively assess structural features.
We find that the nucleonic densities feature slightly lower short-range components compared to the $\Lambda$ density resulting a halo structure in, both, neutrons and protons that extends beyond the $\Lambda$ radius.

\section{Conclusions}

In this work we have transferred the concept of natural orbitals to hypernuclei in an ab initio framework with realistic interactions from chiral effective field theory.
We were able to show that their outstanding properties, i.e., the independence on the HO frequency and the accelerated convergence, can be reproduced in systems with strangeness for ground-state energies and spectra.
To some degree this also holds for radii, however, we have found a remaining frequency dependence in spatially extended hypernuclei inflicted by the truncation of the underlying HO single-particle basis. 

We have further employed the natural orbital wavefunctions together with one-body densities and point-particle radii as a diagnostic tool to study the halo structure of $A=5,6,$ and $7$ hypernuclei.
For \lelem{He}{5} we have found strong indications of a hyperon halo and the addition of further nucleons yields multi-layer halo structures with hyperon and nucleon halos around an $\alpha$ core. 

Finally, we have addressed the occurrence of negative occupation numbers in the NAT basis for open shell systems.
While these can result in pathological behaviors in some systems such as the triply open-shell \lelem{Li}{7}, their overall impact on the many-body calculation is small. 
However, singly-strange hypernuclei are inherently open-shell systems and, thus, the extension to multi-reference treatments for the natural orbitals is a promising lead for future developments. 

Overall, natural orbitals considerably improve the description of p-shell hypernuclei and yield valuable insight into the structure of these systems and with that the hyperon-nucleon interaction. 
Moreover, they pave the way towards ab initio calculations of medium-mass hypernuclei, e.g., within the in-medium NCSM \cite{Gebrerufael2017Ab,Knoell2024Diss}.

\section*{Acknowledgements}

This work is supported by the Deutsche Forschungsgemeinschaft (DFG, German Research Foundation) through the DFG Sonderforschungsbereich SFB 1245 (Project ID 279384907) and the BMBF through Verbundprojekt 05P2024 (ErUM-FSP T07, Contract No. 05P24RDB).
Numerical calculations have been performed on the LICHTENBERG II cluster at the computing center of the TU Darmstadt.

\bibliographystyle{elsarticle-num}
\bibliography{bib_hnucl.bib}

\end{document}